\documentstyle[aps,preprint,epsf]{revtex}
\begin{document}
\draft

\title{Coulomb interaction in oxygen \textit{p}-shell in LDA+U method
and its influence on calculated spectral and magnetic properties of transition metal oxides.}

\author{I.A.~Nekrasov, M.A.~Korotin, V.I.~Anisimov}

\address{Institute of Metal Physics, 
Russian Academy of Sciences-Ural Division, \\
620219 Yekaterinburg GSP-170, Russia}

\date{\today}
\maketitle

\begin{abstract}

Coulomb interaction between electrons on \textit{p}-orbitals
of oxygen atom in strongly correlated compounds is not negligible, since its 
value~($U_p$) has comparable
order of magnitude with the value of Coulomb interaction on \textit{d}-orbitals
of transition metal atom~($U_d$).
We investigate the effect of taking into account Coulomb correlations in 
oxygen \textit{p}-shell in addition to the correlations in the transition metal
\textit{d}-shell in frame of the LDA+U method. 
Our calculations for NiO, MnO and La$_2$CuO$_4$ show that this additional correction
in general improves the agreement with experimental data for the spectral (energy gap values,
relative position of the main peaks in X-ray photoemission spectroscopy (XPS) and Bremsstrahlung 
isohromate spectroscopy (BIS)) and magnetic properties
(magnetic moment values and intersite exchange interaction parameters values).
\end{abstract}


\section{Introduction}

During last several decades the electronic structure calculations 
from first principles became an important part of the solid state
theory. The solution of such essentially 
many-particle problem as
calculation of band structure of real materials is impossible without 
rather severe approximations. The most
famous and commonly used approximation in
\textit{ab~initio} electronic structure
calculations is density functional theory (DFT)~\cite{kohn65} within the local 
spin density approximation
(LSDA). But, as an approximation based on the
homogeneous electron gas theory~\cite{hedin}, the LSDA is 
valid only for compounds with slow
varying through the crystal charge density. In other words, the LSDA 
must describe well only the delocalized electronic
states (broad bands). 
Nevertheless the LSDA is sometimes able to give
correct ground state properties for the systems with rather narrow bands 
(see for example~\cite{Oguchi}).

The most unusual physical properties were found in the systems with strong
electron-electron Coulomb correlations (such as Mott insulators,~ 
high-$T_c$ superconductors,
\textit{etc.}). These systems have been intensively investigated 
during last 20 years both by experimentalist
and theoretician communities. All intriguing features of these systems 
come from the existence of nearly localized
electronic states (narrow bands) such as~\textit{d}~or~\textit{f}  
states of transition metals ions or
rare-earth metals ions, respectively. To mark out the localized state 
one can apply the following criteria:
kinetic energy of localized states is of the same or even smaller order 
of magnitude as the energy of Coulomb interactions.
It is known that for strongly-correlated systems the LSDA often fails 
(high-$T_c$ related compound
La$_2$CuO$_4$; insulating, antiferromagnetic transition metal oxides).
However constrained LSDA calculations~\cite{diderichcs} 
give Coulomb interaction parameter values in surprisingly good
agreement with the experimental estimations 
~\cite{diderichcs,norman,McMahan,Hybertsen,Gunnarsson1,Gunnarsson2,ucalc,Solovyev}.

Several approaches were built on the LSDA basis repairing its 
deficiency in describing Coulomb interaction between localized states.
The most popular methods are the self-interaction correction method (SIC)~\cite{Perdew,sic} 
and the LDA+U method~\cite{ldau1}.

The basic problem of the LSDA is the orbital-independent potential
which does not allow to reproduce Coulomb interaction derived 
energy splitting
between occupied and empty subbands.
The SIC method solves this problem by introducing the orbital-dependent potential
correction which explicitly substracts the self-interaction present
in the LSDA. This method restores correct electronic properties of the transition
metal oxides where the LSDA fails. 
However the self-interaction correction for the \textit{d}~states is so strong, that
when one implements SIC potential only to the
\textit{d}-orbitals of transition metal then the oxygen \textit{p}-orbitals 
do not shift from the LSDA obtained
positions and the occupied \textit{d}-band lies much lower in energy than 
oxygen valence band, which does not agree with the spectroscopy data.
However the values of energy gaps and the spin magnetic moments are in 
rather good agreement with experiment~\cite{sic}. To improve this situation 
one can treat all valence states (namely
the transition metal ions \textit{d}-orbitals and oxygen \textit{p}-orbitals) 
as localized and apply SIC potential to
all of them~\cite{Fujiwara}. In this case structure of occupied bands is 
well reproduced, but the value
of energy gaps will be overestimated~\cite{sicproblem}.
 
Another way to overcome the well known disadvantages of the LSDA is the LDA+U method, 
which gives better agreement with 
experimental spectra~\cite{ldau1}. The LDA+U method corresponds to the static 
limit of recently developed
new many-body approach~---~the dynamical mean-field theory 
(DMFT)~\cite{vollha93}. In its standard form the LDA+U takes
into account only Coulomb interaction between~\textit{d}~(or~\textit{f})~electrons of transition
metal ions. In the present paper we investigate the problem of Coulomb
interaction between oxygen \textit{p}~electrons and show that the inclusion of the
corresponding term in LDA+U equations leads to significant improvement of 
agreement between calculated and experimental spectral and magnetic properties.

\section{Method of calculation}

The main idea of the LDA+U method is to add to
the LSDA functional the term $E^U$ corresponding to the mean-field approximation
of the Coulomb interaction in multiband Hubbard model.
\begin{equation}  \label{U1}
E^{LDA+U}[\rho ^\sigma ({\bf r}),\{n^\sigma \}]=E^{LSDA}[\rho ^\sigma ({\bf %
r)}]+E^U[\{n^\sigma \}]-E_{dc}[\{n^\sigma \}],
\end{equation}
where $\rho ^\sigma ({\bf r})$ is the charge density for spin-$\sigma $
electrons and $E^{LSDA}[\rho ^\sigma ({\bf r})]$ is the standard LSDA
(Local Spin-Density Approximation) functional. Eq.~(\ref{U1}) asserts that the LSDA
is sufficient in the absence of orbital polarizations, while the latter are
driven by, 
\begin{equation}  \label{upart}
\begin{array}{c}
E^U[\{n^\sigma\}]=\frac 12\sum_{\{m\},\sigma }\{\langle m,m^{\prime \prime }\mid
V_{ee}\mid m^{\prime },m^{\prime \prime \prime }\rangle n_{mm^{\prime
}}^\sigma n_{m^{\prime \prime }m^{\prime \prime \prime }}^{-\sigma }+ \\ 
\\ 
(\langle m,m^{\prime \prime }\mid V_{ee}\mid m^{\prime },m^{\prime \prime
\prime }\rangle -\langle m,m^{\prime \prime }\mid V_{ee}\mid m^{\prime
\prime \prime },m^{\prime }\rangle )n_{mm^{\prime }}^\sigma n_{m^{\prime
\prime }m^{\prime \prime \prime }}^\sigma \},
\end{array}
\end{equation}
where~$V_{ee}$ are the screened Coulomb interactions among the \textit{d}~electrons.
Finally, the last term in Eq.~(\ref{U1}) corrects for double
counting (in the absence of orbital polarizations, Eq.~(\ref{U1}) should
reduce to~$E^{LSDA}$) and is given by

\begin{equation}  \label{U3}
E_{dc}[\{n^\sigma \}]=\frac 12UN(N-1)-\frac 12J[N^{\uparrow }(N^{\uparrow
}-1)+N^{\downarrow }(N^{\downarrow }-1)],
\end{equation}
were $N^\sigma =Tr(n_{mm^{\prime }}^\sigma )$ and $N=N^{\uparrow
}+N^{\downarrow }$. $U$~and~$J$ are screened Coulomb and exchange parameters 
~\cite{Gunnarsson1,ucalc}.

In addition to the usual LSDA potential, an effective single-particle
potential to be used in the effective single-particle Hamiltonian has the form:

\begin{equation}  \label{hamilt}
\widehat{H}=\widehat{H}_{LSDA}+\sum_{mm^{\prime }}\mid inlm\sigma \rangle
V_{mm^{\prime }}^\sigma \langle inlm^{\prime }\sigma \mid
\end{equation}
\begin{equation}  \label{Pot}
\begin{array}{c}
V_{mm^{\prime }}^\sigma =\sum_{m^{\prime \prime}m^{\prime \prime \prime}}
\{\langle m,m^{\prime \prime }\mid
V_{ee}\mid m^{\prime },m^{\prime \prime \prime }\rangle n_{m^{\prime \prime
}m^{\prime \prime \prime }}^{-\sigma }+ \\ 
\\ 
(\langle m,m^{\prime \prime }\mid V_{ee}\mid m^{\prime },m^{\prime \prime
\prime }\rangle -\langle m,m^{\prime \prime }\mid V_{ee}\mid m^{\prime
\prime \prime },m^{\prime }\rangle )n_{m^{\prime \prime }m^{\prime \prime
\prime }}^\sigma \}- \\ 
\\ 
U(N-\frac 12)+J(N^{\sigma}-\frac 12).
\end{array}
\end{equation}
The matrix elements of Coulomb interaction can be expressed in terms of complex spherical harmonics and
effective Slater integrals~$F^k$~\cite{JUDD} as 
\begin{equation}  \label{slater}
\langle m,m^{\prime \prime }\mid V_{ee}\mid m^{\prime },m^{\prime \prime
\prime }\rangle =\sum_ka_k(m,m^{\prime },m^{\prime \prime },m^{\prime \prime
\prime })F^k,
\end{equation}
where~$0\leq k\leq 2l$ and

\[
a_k(m,m^{\prime },m^{\prime \prime },m^{\prime \prime \prime })=\frac{4\pi }{%
2k+1}\sum_{q=-k}^k\langle lm\mid Y_{kq}\mid lm^{\prime }\rangle \langle
lm^{\prime \prime }\mid Y_{kq}^{*}\mid lm^{\prime \prime \prime }\rangle 
\]
For~\textit{d}~electrons one needs~$F^0,F^2$ and~$F^4$ and these can be linked to the
Coulomb- and Stoner parameters~$U$ and~$J$ obtained from the LSDA-supercell
procedures via~$U=F^0$ and~$J=(F^2+F^4)/14$, while the ratio~$F^2/F^4$ is to
a good accuracy a constant~$\sim 0.625$ for the~3\textit{d} elements~\cite{deGroot,ANISOL}.
(For~\textit{f}~electrons the corresponding expression is~$J=(286F^2+195F^4+250F^6)/6435$).
The Coulomb parameter~$U$ is calculated as
a second derivative of the total energy (or the first derivative of the
corresponding eigenvalue)  in respect to the occupancy of localized orbitals
of the central atom in a supercell with fixed occupancies on all other
atoms~\cite{Gunnarsson1}.

If one neglects the exchange and non-sphericity of the Coulomb interaction
(which is exact in the case of the fully occupied or empty band)
the potential correction will have the more simple form:
\begin{equation}  \label{vlda}
V_i=U(\frac 12-n_i)\;
\end{equation}
where $n_i$ is the occupancy of $i$-orbital. Then for fully occupied state
LDA+U potential correction would be the shift to the lower energies on 
$U/2$, while for empty states it gives an upward shift on the same
value. So the LDA+U gives correct splitting between occupied and empty
subbands equal to the Coulomb interaction parameter~$U$.

In the LDA+U approach the Coulomb interactions are taken into 
account conventionally only on
\textit{d}-orbitals of transition metals. However it is known that 
Coulomb interactions between
electrons on \textit{p}-orbitals of oxygen have comparable order of magnitude
~\cite{McMahan,Hybertsen} with the corresponding~$d-d$ Coulomb interactions and so must be 
taken into consideration on the same footing as for \textit{d}-orbitals. 
The usual justification for omitting of $U$ on oxygen \textit{p}-shell is that the oxygen shell is fully
occupied and the correlation effects between electrons (or rather holes)
in it can be neglected due to the small number of holes in ground state.
However the  LDA+U equations~(\ref{vlda}) will give nonzero correction
for the fully occupied oxygen band:
\begin{equation}
V_p=-U_p/2
\end{equation}
This potential correction must be applied to the orbitals forming oxygen band,
however corresponding Wannier functions (in contrast to \textit{d}~states) 
are far from being of pure O(2\textit{p}) character
because they have very strong admixture of~\textit{s}~and~\textit{p}~states of transition metal
ions and other extended orbitals. Since the main influence on the electronic structure
is the change of the energy separation between the oxygen \textit{p}-band and the
transition metal \textit{d}-band,
the upward shift in energy of the transition metal \textit{d}-band on~$U_p/2$ will 
be equivalent to the shifting down of the oxygen \textit{p}-band on the same value.
Thus in our calculations we
added~$U_p/2$ term to the diagonal matrix elements of the LDA+U potential correction~(\ref{Pot}).
We call this extension of the LDA+U method in
the paper as the~LDA+U$^{(d+p)}$.

Recently the modified~LDA+U$^{(d+p)}$ method was used 
by Korotin~\textit{et al.} for investigation of charge and orbital ordering
effects in La$_{7/8}$Sr$_{1/8}$MnO$_3$ compound~\cite{Korotin}. The inclusion of
Coulomb interactions in
the oxygen \textit{p}-shell was found to be crucial in that calculation, since it 
controls the value
of charge transfer energy between Mn(3\textit{d}) and O(2\textit{p}) valence 
states and significantly enhances the tendency of localization in this
system.

In the present paper we report the results with the use of the modified LDA+U$^{(d+p)}$ method
for the typical strongly correlated transition metal oxides NiO, MnO and La$_2$CuO$_4$.
We show that inclusion of the correlations in oxygen \textit{p}-shell
leads to the better agreement with the experimental data for 
the main peaks position in
X-ray photoemission spectroscopy (XPS) and Bremsstrahlung 
isohromate spectroscopy (BIS) spectra in comparison with conventional LDA+U calculated spectra.
Not only spectral properties, but both spin magnetic moments and intersite exchange 
interaction parameters~$J_{ex}$
for NiO, MnO and~La$_2$CuO$_4$ are in better agreement with the corresponding 
experimental data.

\section{Results and discussion}

The important part of the LDA+U calculation scheme is the determination of Coulomb
interaction parameters~$U$~and~$J$ in equations~(\ref{Pot}): 
Coulomb parameter~$U_p$ for
\textit{p}-orbitals of oxygen,~$U_d$ for transition metals ion and Hund's parameter~$J$ for
\textit{d}-orbitals of transition metals.
To get~$U_d$~and~$J$ one can use the supercell procedure~\cite{ucalc,Solovyev} or
the constrained LSDA method~\cite{Gunnarsson1}, which are
based on calculation of the variation of the total energy as a function of the local occupation of the 
\textit{d}-shell. We took the values of~$U_d$ and~$J$ parameters 
listed in Table~I previously calculated~in~\cite{ldau1}. The problem is
how to determine the Coulomb parameter~$U_p$.

Due to more extended nature of the~O(2\textit{p}) Wannier states in comparison with 
transition metal \textit{d}~states, the constrained occupation calculations can not
be implemented as easy as for the~\textit{d}-shell of transition metals.
Nevertheless several independent and different techniques were used for this purpose previously by
different authors. McMahan~\textit{et al.} estimated the value of~$U_p$ in high-$T_c$ related
compound~La$_2$CuO$_4$ using the constrained LDA calculation 
where only atomic-like O(2\textit{p})-orbitals within oxygen atomic spheres were considered
instead of the more extended Wannier functions. The corresponding value of Coulomb interaction
parameter~$U_p$ was obtained as~$7.3~eV$. This value can be considered as the upper limit
of the exact~$U_p$. The LDA calculations gave the estimation that
only~75\% of Wannier function density lies in the oxygen atomic sphere so that
renormalized value of Coulomb interaction parameter for oxygen Wannier functions
is~$U_p=(7.3) \times (0.75)^2=4.1~eV$~\cite{McMahan}.

Later Hybertsen~\textit{et al.}
suggested the scheme to calculate~$U_p$, which consists of two steps: (i)~via 
constrained-density-functional
approach one can obtain the energy surface~$E(N_d,N_p)$ as a function of local charge 
states and (ii)~simultaneously extended Hubbard model was solved in mean-field approximation
as a function of local charge states~$N_d$ and~$N_p$.
Corresponding Coulomb interaction parameters were extracted as those which give the
energy surface matching the microscopic density-functional calculations 
results~\cite{Hybertsen}. The obtained values for~$U_p$ are~$3 \div 8~eV$
depending on the parameters of calculations. 

Another way to estimate~$U_p$ is to use Auger spectroscopy data, where two
holes in O(2\textit{p})-shell are created in the excitation process. Such fitting to the experimental
spectra gave the value of~$U_p=5.9~eV$~\cite{Knotek}.
In our~LDA+U$^{(d+p)}$ calculations we used~$U_p=6~eV$. 

To solve the LDA+U hamiltonian we implemented the
self-consistent tight-binding~(TB) linear muffin-tin orbitals method~(LMTO) in the
atomic sphere approximation~(ASA)~\cite{Andersen1,Andersen2,calcfit}. For the calculations
we choose the classical strongly correlated transition metal oxides NiO, MnO and 
La$_2$CuO$_4$, which were well investigated by experimental and theoretical methods.

Comparison between the LDA+U (left column) and the LDA+U$^{(d+p)}$ (right column) calculated 
density of states (DOS)
of NiO, MnO and La$_2$CuO$_4$ is presented in 
figures~1,~2~and~3. For all compounds one can 
see that the main difference between the LDA+U$^{(d+p)}$ and the LDA+U calculated densities 
of states is the increased energy separation between the oxygen~2\textit{p} and transition
metal~3\textit{d} bands. The larger value of "charge transfer" energy~(O(2\textit{p})-Me(3\textit{d}))
(Me=Ni,Mn,Cu) leads to the enhanced ionicity and decreased covalency nature of the electronic structure:
the unoccupied bands have more pronounced~3\textit{d} character and the admixture of
oxygen states to those bands becomes weaker.

The ground state is correctly described both by LDA+U and 
LDA+U$^{(d+p)}$ calculations as antiferromagnetic insulator for all compounds. 
The values of
energy gaps~\cite{gap} and spin magnetic moments are presented in 
tables~II~and~III. One can see that
the values obtained in the LDA+U$^{(d+p)}$ calculations
are in general in better agreement with experiment than the LDA+U calculated values.
While the increasing of the energy gap values with applying~$U_p$ correction
was obviously expected with the increasing of "charge transfer" energy 
in the compounds belonging to the class of "charge transfer" insulators~\cite{zsa},
the increasing of the magnetic moments values is more complicated self-consistency
effect due to the increased ionicity in the LDA+U$^{(d+p)}$ calculations comparing with
the LDA+U results.

In Fig.~4 the DOS obtained by LDA+U$^{(d+p)}$ and LDA+U calculations
for MnO and NiO compounds
are compared with the superimposed XPS and BIS spectra corresponding to the removal
of an electron (the occupied
bands) and addition of an electron (the empty bands), respectively. The better agreement
with the experimental data
of position of the main peaks of unoccupied band relative to the occupied one
is the direct confirmation of the importance of taking into account Coulomb
interactions in oxygen~2\textit{p}-shell.

The values of the intersite exchange
interaction parameters~$J_{ex}$ depend on the parameters of the electronic structure
in a rather indirect implicit way. The developing of the good calculating scheme
for exchange parameters is very important because the \textit{ab-initio} calculation
is often the only way to describe the magnetic properties of complicated compounds
such as for example "spin-gap" systems~\cite{spin-gap}. Recently Solovyev \textit{et al.}~\cite{sol}
did very through analysis of the exchange interaction parameters for MnO
calculated using different methods of  electronic structure calculations.
They used the positions of the~Mn(3\textit{d})-spin-up and~Mn(3\textit{d})-spin-down bands relative to the
oxygen~2\textit{p} states as adjustable parameters to fit the values of exchange interaction
for the nearest and second~Mn-Mn neighbors. Their results gave nearly the same
splitting between~Mn(3\textit{d})-spin-up and~Mn(3\textit{d})-spin-down states as in standard
LDA+U calculations~
($10.6~eV$) but the position of those states relative to the oxygen band 
was shifted approximately on~$3~eV$ up relative to the LDA+U case. It is practically the same as
we have in our LDA+U$^{(d+p)}$ calculations, because with~$U_p=6~eV$ the shift
of the position~Me(3\textit{d})-band relative to the oxygen~O(2\textit{p})-band is equal to~$U_p/2=3~eV$.

Comparison between LDA+U and LDA+U$^{(d+p)}$ calculated~$J_{ex}$ parameters and 
experimental data is presented in table~IV.
$J_{ex}$ were calculated from Greens function method as second 
derivatives of the
ground state energy with respect to the magnetic moment rotation angle~\cite{Lichtenstein,obmeny}.
Again one can see that in general the LDA+U$^{(d+p)}$ gives better results than
the LDA+U, especially for MnO compound.
 
\section{Conclusion}

The method for inclusion of Coulomb interactions between oxygen~\textit{p}~electrons
in the calculation scheme of the LDA+U method was proposed. The main effect
was found to be the increasing of "charge transfer" energy parameter
(the separation of~O(2\textit{p}) and~Me(3\textit{d}) states). As the result, the spectral and magnetic
properties of the typical strongly correlated transition metal oxides NiO, MnO and La$_2$CuO$_4$
were found in better agreement with experimental data than in the conventional
LDA+U method where only correlations between~Me(3\textit{d}) state are taken into account.

\acknowledgements
We are greatful to T. Fujiwara for the helpful discussions.
This work was supported by the Russian Foundation for Basic Research (RFFI-98-02-17275).

\begin{figure}
\label{lacuofig}
\vskip2mm
\epsfxsize=10cm
\centerline{ \epsffile{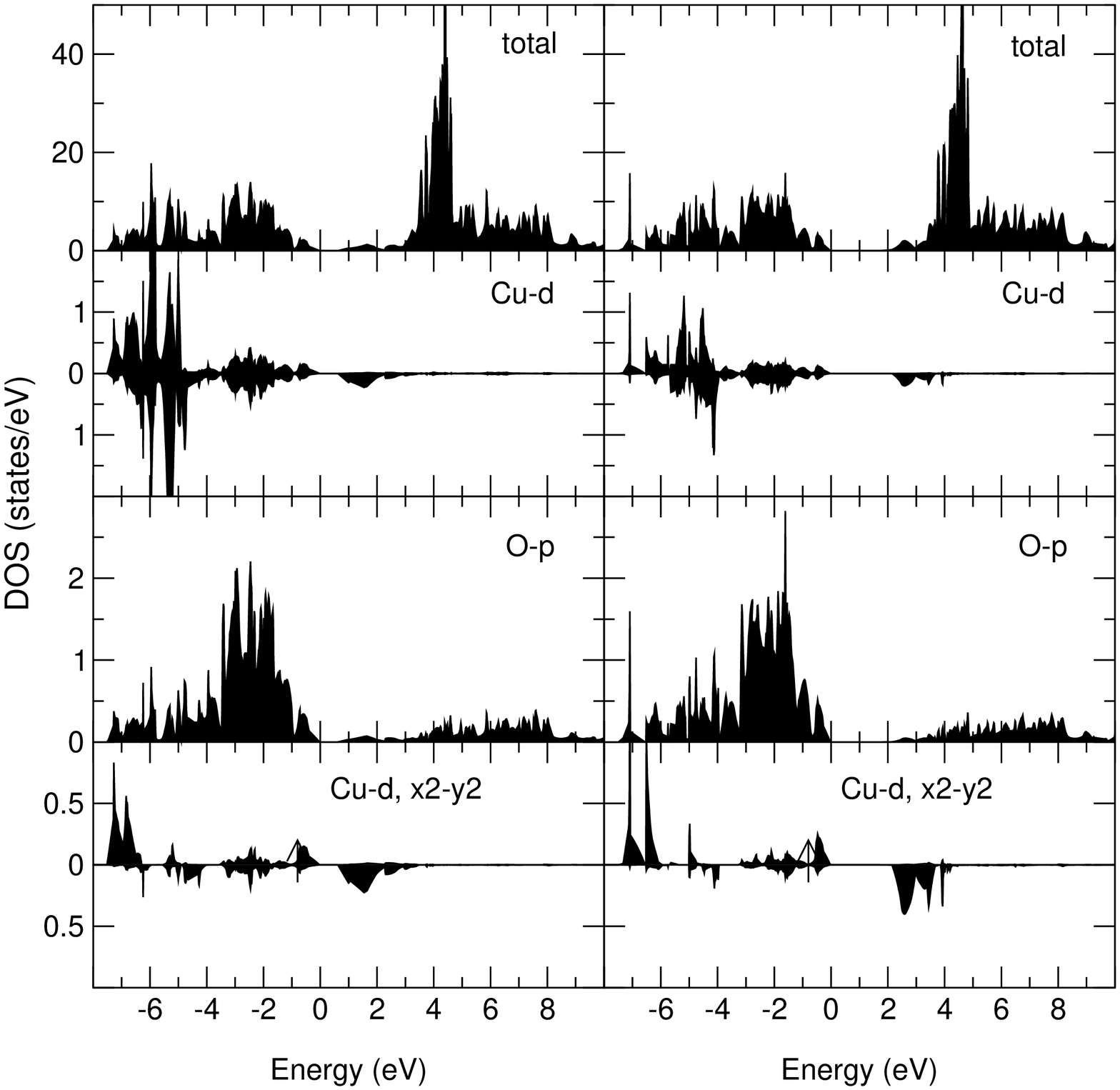} }

\narrowtext
\caption {La$_2$CuO$_4$ DOS calculated by the LDA+U~(left column) and the LDA+U$^{(d+p)}$~(right column)
methods. (On all figures the total DOS is presented per formula
unit, the DOS of particular states are per atom. Fermi energy corresponds to zero.)}

\end{figure}

\begin{figure}
\label{mnofig}
\vskip2mm
\epsfxsize=10cm
\centerline{ \epsffile{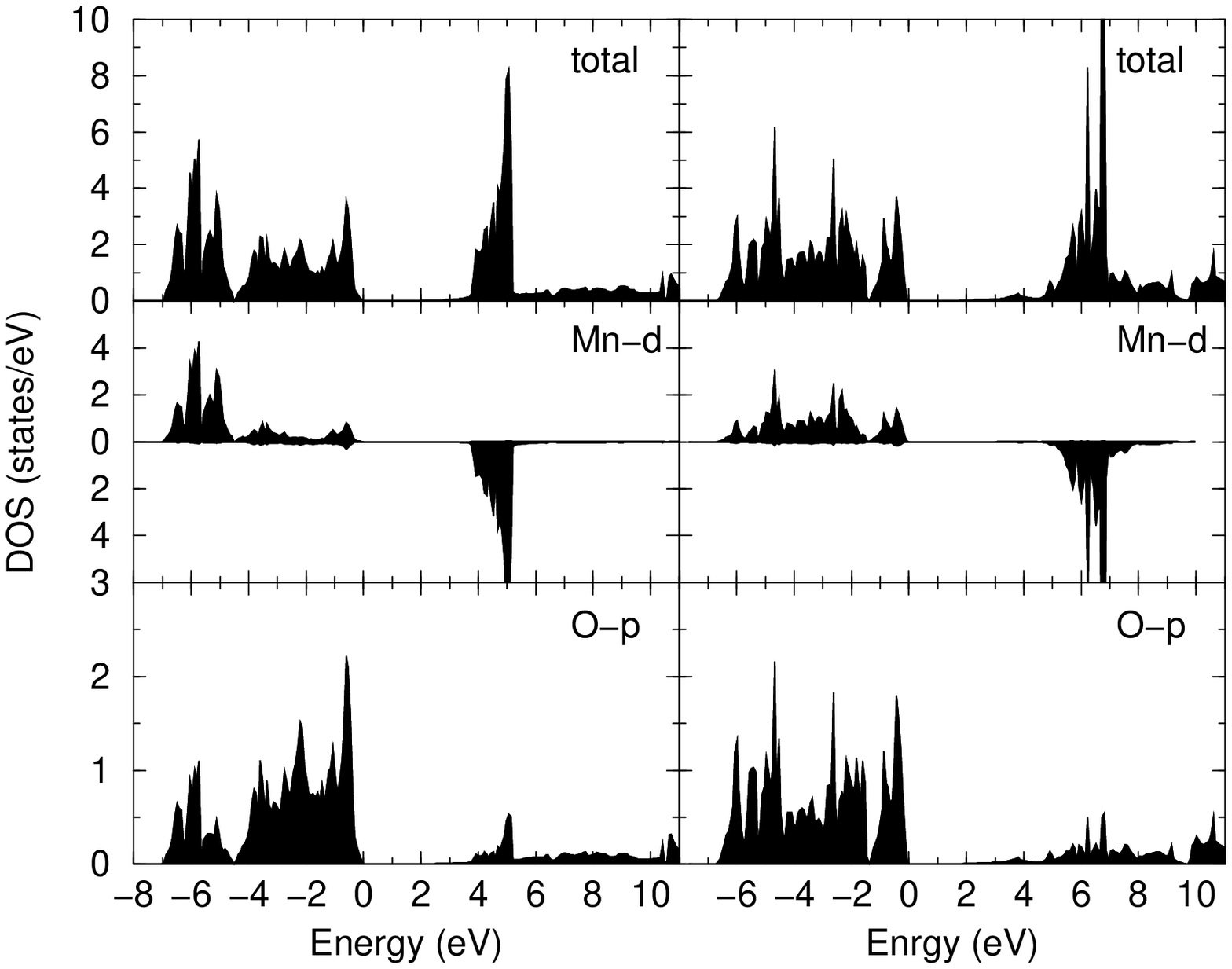} }

\narrowtext
\caption {MnO DOS calculated by the LDA+U~(left column) and the LDA+U$^{(d+p)}$~(right column) methods.}

\end{figure}

\begin{figure}
\label{niofig}
\vskip2mm
\epsfxsize=10cm
\centerline{ \epsffile{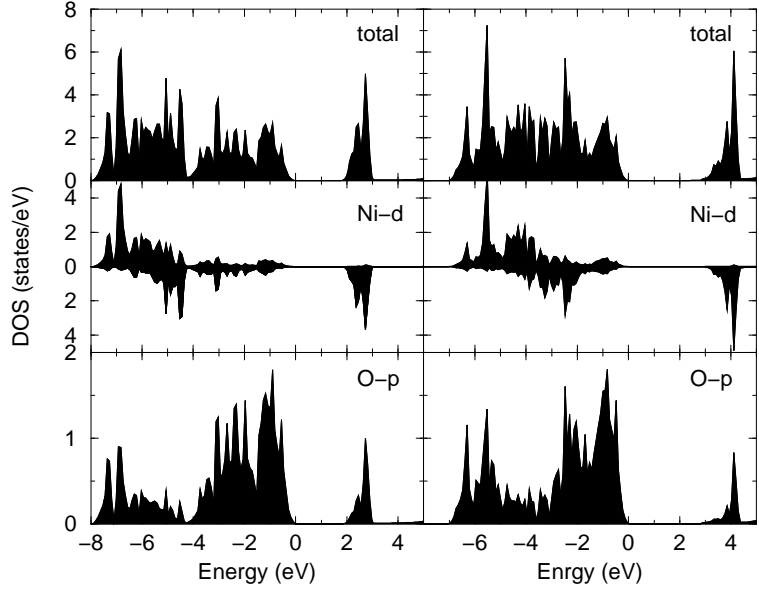} }

\narrowtext
\caption {NiO DOS calculated by the LDA+U~(left column) and the LDA+U$^{(d+p)}$~(right column) methods.}

\end{figure}

\begin{figure}
\label{mnonio}
\vskip2mm
\epsfxsize=10cm
\centerline{ \epsffile{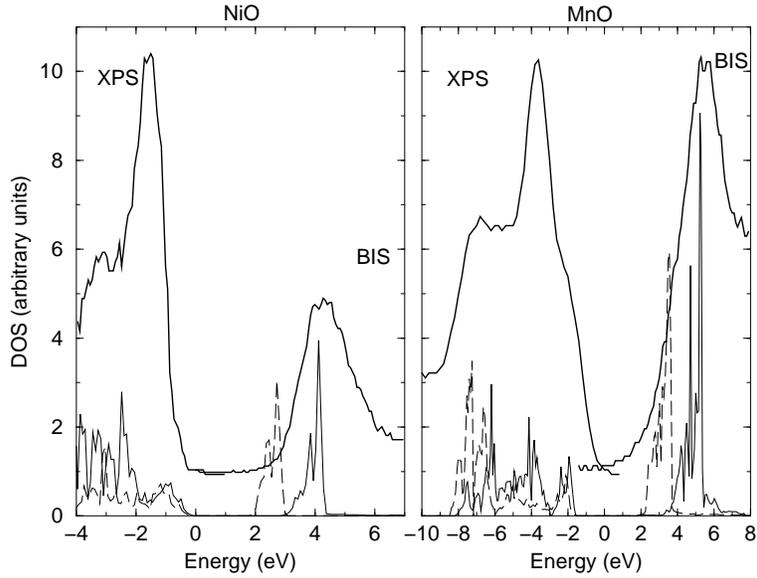} }

\narrowtext
\caption {DOS calculated by the LDA+U (dashed line) and the LDA+U$^{(d+p)}$ (solid line)~Ni(3\textit{d})
and~Mn(3\textit{d}) in comparison with superimposed XPS and BIS spectra.}
Spectroscopic data:\\
MnO:~G.A. Sawatzky and J.W. Allen, Phys. Rev. Lett. \textbf{53}, 2339 (1984).\\
NiO:~J. van Elp, R. H. Potze, H. Eskes, R. Berger, and G. A. Sawatzky, Phys.Rev.B \textbf{44}, 1530 (1991).
\end{figure}

\begin{table}
\label{parameters}

\caption{Coulomb parameters~$U_d$ and Hund's parameters~$J$~($eV$) used in calculations.}

\begin{tabular}{cccccc}
\multicolumn{1}{c}{ } &
\multicolumn{1}{c}{La$_2$CuO$_4$} &
\multicolumn{1}{c}{MnO} &
\multicolumn{1}{c}{NiO} \\
\hline
\multicolumn{1}{l}{$U_d$}&
\multicolumn{1}{c}{8.0} &
\multicolumn{1}{c}{6.9} &
\multicolumn{1}{l}{8.0} \\
\multicolumn{1}{l}{$J$}&
\multicolumn{1}{c}{1.0} &
\multicolumn{1}{c}{0.86} &
\multicolumn{1}{l}{0.95} \\
\end{tabular}

\end{table}

\begin{table}
\label{gap}

\caption{Calculated and experimental values of energy gaps ($eV$).}

\begin{tabular}{cccccc}
\multicolumn{1}{c}{ } &
\multicolumn{1}{c}{LDA+U} &
\multicolumn{1}{c}{LDA+U$^{(d+p)}$} &
\multicolumn{1}{c}{Experiment} \\
\hline
\multicolumn{1}{l}{La$_2$CuO$_4$}&
\multicolumn{1}{c}{0.7} &
\multicolumn{1}{c}{2.0} &
\multicolumn{1}{l}{2.0$^a$ } \\
\multicolumn{1}{l}{MnO}&
\multicolumn{1}{c}{3.8} &
\multicolumn{1}{c}{4.5} &
\multicolumn{1}{l}{$3.6-3.8^b$ } \\
\multicolumn{1}{l}{NiO}&
\multicolumn{1}{c}{1.8} &
\multicolumn{1}{c}{2.8} &
\multicolumn{1}{l}{$4.3^c,4.0^d$ } \\
\end{tabular}

$^a$S. Uchida, {\it et al.}, Phys. Rev. B {\bf 43}, 7942 (1991). \\
$^b$L. Messick, {\it et al.}, Phys. Rev. B {\bf 6}, 3941 (1972). \\
$^c$G.A. Sawatzki and J.V. Allen, Phys. Rev. Lett. {\bf 53}, 2329 (1984). \\
$^d$S. H\"{u}fner, {\it et al.}, Solid State Commun. {\bf 52}, 793 (1984). \\

\end{table}

\begin{table}
\label{moments}

\caption{Calculated and experimental values of spin magnetic moments ($\mu_B$).}

\begin{tabular}{cccccc}
\multicolumn{1}{c}{ } &
\multicolumn{1}{c}{LDA+U} &
\multicolumn{1}{c}{LDA+U$^{(d+p)}$} &
\multicolumn{1}{c}{Experiment} \\
\hline
\multicolumn{1}{l}{La$_2$CuO$_4$}&
\multicolumn{1}{c}{0.45} &
\multicolumn{1}{c}{0.68} &
\multicolumn{1}{l}{$0.60^a$ } \\
\multicolumn{1}{l}{MnO}&
\multicolumn{1}{c}{4.51} &
\multicolumn{1}{c}{4.59} &
\multicolumn{1}{l}{$4.79^b,4.58^c$ } \\
\multicolumn{1}{l}{NiO}&
\multicolumn{1}{c}{1.50} &
\multicolumn{1}{c}{1.64} &
\multicolumn{1}{l}{$1.77^d, 1.64^e, 1.90^f$ } \\
\end{tabular}

$^a$Y. Endoh, {\it et al.}, Phys. Rev. B {\bf 37}, 7443 (1988). \\
$^b$F.P. Koffyber and F.A. Benko, J. Appl. Phys. {\bf 53}, 1173 (1982). \\
$^c$J.B. Forsith, {\it et al.}, J. Phys. {\bf 21}, 2917 (1988). \\
$^d$O.K. Andersen and O. Jepsen, Phys. Rev. Lett. {\bf 53}, 2571 (1984). \\
$^e$H.A. Alperin, J. Phys. Soc. Jpn. Suppl. B {\bf 17}, 12 (1962). \\
$^f$A.K. Cheetham and D.A.O. Hope, Phys. Rev. B {\bf 27}, 6964 (1983). \\

\end{table}

\begin{table}
\label{exchange}

\caption{Calculated and experimental values of intersite exchange interaction parameters~$J_{ex}$~($meV$).}

\begin{tabular}{cccccc}
\multicolumn{1}{c}{ } &
\multicolumn{1}{c}{LDA+U} &
\multicolumn{1}{c}{LDA+U$^{(d+p)}$} &
\multicolumn{1}{c}{Experiment} \\
\hline
\multicolumn{1}{l}{$^1$La$_2$CuO$_4$}&
\multicolumn{1}{c}{82.9} &
\multicolumn{1}{c}{100.9} &
\multicolumn{1}{c}{$136\pm5^a$} \\
\multicolumn{1}{l}{$^2$MnO}&
\multicolumn{1}{c}{5.4~~~9.3} &
\multicolumn{1}{c}{5.4~~~5.1} &
\multicolumn{1}{c}{$4.8^b,5.4^c~~~5.6^b,5.9^c$} \\
\multicolumn{1}{l}{$^2$NiO}&
\multicolumn{1}{c}{0.8~~~23.2} &
\multicolumn{1}{c}{0.2~~~19.4} &
\multicolumn{1}{c}{$1.4^d~~~~19.0^d$} \\
\end{tabular}

$^1$Cu-Cu exchange parameter between nearest Cu atoms in plane.\\
$^2$Me-Me exchange parameters between nearest and second neighbors.\\
$^a$G. Aeppli, {\it et al.}, Phys. Rev. Lett. {\bf 62}, 2052 (1989). \\
$^b$M. Kohgi, {\it et al.}, Solid State Commun. {\bf 11}, 391 (1972). \\
$^c$M.E. Lines and E.D. Jones, Phys. Rev. {\bf 139}, A1313 (1965). \\
$^d$M.I. Hutchings and S.J. Samuelson, Solid State Commun. {\bf 9}, 1011 (1971). \\
\end{table}


\begin{thebibliography}{99}

\bibitem{kohn65}W. Kohn, L.J. Sham, Phys. Rev. A - Gen.Phys. \textbf{140}, 1133 (1965);
L.J. Sham, W. Kohn, Phys. Rev. \textbf{145 N 2}, 561 (1966);

\bibitem{hedin}L. Hedin and B. Lundqvist J. Phys. C: Solid State Phys. \textbf{4}, 2064 (1971);
U. von Barth and L. Hedin J. Phys. C: Solid State Phys. \textbf{5}, 1629 (1972);

\bibitem{Oguchi}K. Terakura, T. Oguchi, A.R. Williams, and J. K\"{u}bler,
Phys. Rev. B \textbf{30}, 4734 (1984).

\bibitem{diderichcs}P.H. Diderchs, S. Bl\"{u}gel, R. Zeller, and H. Akai,
Phys. Rev. Lett. \textbf{53}, 2512 (1984).

\bibitem{norman}M.R. Norman and A.J. Freeman,
Phys. Rev. B \textbf{33}, 8896 (1986).

\bibitem{McMahan}A.K. McMahan, R.M. Martin, and S. Satpathy,
Phys. Rev. B \textbf{38}, 6650 (1988).

\bibitem{Hybertsen}M.S. Hybertsen, M. Schl\"{u}ter, and N.E. Christensen,
Phys. Rev. B \textbf{39}, 9028 (1989).

\bibitem{Gunnarsson1}O. Gunnarsson, O. K. Andersen, O. Jepsen, and J. Zaanen,
Phys. Rev. B \textbf{39}, 1708 (1989).

\bibitem{Gunnarsson2}O. Gunnarsson, A.V. Postnikov, and O.K. Andersen,
Phys. Rev. B \textbf{40}, 10407 (1989).

\bibitem{ucalc}V.I. Anisimov, and O. Gunnarsson,
Phys. Rev. B \textbf{43}, 7570 (1991).

\bibitem{Solovyev}I.V. Solovyev, P.H. Dederichs, and V.I. Anisimov,
Phys. Rev. B \textbf{50}, 16861 (1994).

\bibitem{Perdew}J.P. Perdew and A. Zunger,
Phys. Rev. B \textbf{23}, 5048 (1981).

\bibitem{sic}A. Svane and O. Gunnarsson,
Phys. Rev. Lett \textbf{65}, 1148 (1990).

\bibitem{ldau1}V.I. Anisimov, J. Zaanen, and O.K. Andersen,
Phys. Rev. B \textbf{44}, 943 (1991).

\bibitem{ldau2}V.I. Anisimov, F. Aryasetiawan, and A.I. Lichtenstein,
J. Phys.: Condens Matter \textbf{9}, 767 (1997). 

\bibitem{Fujiwara}M. Arai and T. Fujiwara,
Phys. Rev. B \textbf{51}, 1477 (1995).

\bibitem{sicproblem}T. Fujiwara, M. Arai, and Y. Ishii in {\em Strong coulomb correlations in
electronic structure calculations: Beyond the local
density approximation}, \textbf{Volume 1}, edited by V.I. Anisimov, Gordon and Breach Science
Publishers, Singapore, 2000, p.~167.

\bibitem{vollha93}D.~Vollhardt in {\em Correlated Electron Systems},
edited by V.J. Emery, World Scientific, Singapore, 1993, p.~57;
Th. Pruschke, M. Jarrell, and J.K. Freericks,
Adv. in Phys. {\bf 44}, 187 (1995);
A. Georges, G. Kotliar, W. Krauth, and M.J. Rozenberg,
Rev. Mod. Phys. {\bf 68}, 13 (1996).

\bibitem{JUDD}B.R. Judd, 'Operator techniques in atomic spectroscopy',
McGrow-Hill, New York, 1963.

\bibitem{deGroot}F.M.F. de Groot, J.C. Fuggle, B.T. Thole, G.A. Sawatzky,
Phys. Rev. B {\bf 42}, 5459 (1990).

\bibitem{ANISOL}V.I. Anisimov, I.V. Solovyev, M.A. Korotin, M.T. Czyzyk, and G.A.Sawatzky,
Phys. Rev. B {\bf 48}, 16929 (1993).

\bibitem{Korotin}M. Korotin, T. Fujiwara, and V. Anisimov,
Phys. Rev. B \textbf{62}, 5696 (2000).

\bibitem{Knotek}M.L. Knotek and P.J. Feibelman,
Phys. Rev. Lett. \textbf{40}, 964 (1978).

\bibitem{Andersen1}O.K. Andersen, Z. Pawlowska, and O. Jepsen,
Phys. Rev. B \textbf{34}, 5253 (1986).

\bibitem{Andersen2}O.K. Andersen, C. Arccangeli, R.W. Tank, T. Saha-Dasgupta, G. Krier, O.Jepsen, and
I. Dasgupta, cond-mat/9804166. 

\bibitem{calcfit}All calculations were done without downfolding, within the orbital basis of~
Me(4\textit{s},4\textit{p},3\textit{d}) (Me=Ni,Mn,Cu) and~O(3\textit{s},2\textit{p},3\textit{d}).
Logarithmic derivatives of~O(3\textit{s}),~O(3\textit{d}) and~Me(3\textit{p}) bands were fixed to make~
$E_\nu$ and~$c$ parameters of the LMTO method to be equal.

\bibitem{gap}One should say
that we look at the energy gap between the highest edge of occupied and the
lowest edge of empty part of~Me(3\textit{d})-band (Me=Ni,Mn,Cu), because
in total DOS the value of energy gap is~$\sim~1~eV$ due to very low intensive~O(3\textit{s})-band.

\bibitem{zsa}J. Zaanen, G.A. Sawatzky, J.W. Allen,
Phys. Rev. Lett. {\bf 55}, 418 (1985).

\bibitem{spin-gap}M.A. Korotin, I.S. Elfimov, V.I. Anisimov, M. Troyer, and D.I. Khpmskii,
Phys. Rev. Lett. {\bf 83}, 1387 (1999).

\bibitem{sol}I.V. Solovyev, K. Terakura,
Phys. Rev. B \textbf{58}, 15496 (1998).

\bibitem{Lichtenstein}A.I. Lichtenstein, M.I. Katsnelson, V.P. Antropov, and V.A. Gubanov,
J. Magn. Magn. Matter \textbf{67}, 65 (1987).

\bibitem{obmeny}To calculate~$J_{ex}$ in MnO and NiO we used the 8-fold enlarged supercell in order to
get the exchange parameters between nearest and second order neighbors. For La$_2$CuO$_4$ the 2-fold
enlarged supercell was used.

\end{thebibliography}
\end{document}